\documentstyle[aps,pre,eqsecnum,preprint]{revtex}

\begin{document}
\bibliographystyle{unsrt}
\textwidth 800pt

\large
\begin{center}
\underline{Perturbation analysis of weakly discrete kinks
}  
\vspace{2cm}\\ \large S. \vspace{0.5cm}Flach$^*$ and K. Kladko\\
\normalsize
Max-Planck-Institut f\"ur Physik komplexer Systeme \\
Noethnitzer Str. 38, D-01187 Dresden, Germany \vspace{1cm} 
\\
\end{center}
\normalsize
ABSTRACT \\
We present a perturbation theory of kink solutions of discrete
Klein-Gordon chains. The unperturbed solutions correspond to
the kinks of the adjoint partial differential equation. 
The perturbation theory is based on  a reformulation 
of the discrete chain problem into a
partial differential equation with spatially modulated  mass density.
The first
order corrections to the kink solutions 
are obtained analytically and are shown to 
agree with exact numerical results.
We discuss the problem of calculating the
Peierls-Nabarro barrier. 
\vspace{0.5cm}
\newline
PACS number(s): 03.20.+i ; 03.40.-t ; 63.20.Ry 
\newline
\newline
\\
\\
\\
\\
\\
$^*$ email: flach@idefix.mpipks-dresden.mpg.de

\newpage

\section{Introduction}

In recent years there has been considerable effort in understanding
the effects of discreteness on soliton-like solutions
\cite{pr82},\cite{im82},\cite{cy83},\cite{vlp81},\cite{twk80},\cite{mr86},\cite{wes86},\cite{sweb86},\cite{wb90},\cite{tm92}. In this work
we will restrict ourselves to kink solutions. Kinks connect two
groundstates of a choosen system. Let us
consider a nonlinear Klein-Gordon equation
\begin{equation}
\frac{\partial^2 \Phi}{\partial t^2} - C\frac{\partial ^2 \Phi}{
\partial x^2} + \frac{\partial V}{\partial \Phi} = 0\;\;.
\label{1-1}
\end{equation}
To allow for kink solutions the potential $V(z)$ has to have at
least two degenerate minima. Throughout this paper we will consider
only static solutions, i.e. the field $\Phi$
will be not time-dependent.
Then equation (\ref{1-1}) is reduced to an ordinary differential equation
\begin{equation}
- C\frac{\partial ^2 \Phi}{
\partial x^2} + \frac{\partial V}{\partial \Phi} = 0\;\;.
\label{1-2}     
\end{equation}
The phase space of (\ref{1-2}) is two-dimensional. A kink solution
corresponds to a heteroclinic orbit. This orbit connects the two
hyperbolic fixed points (the groundstates) in phase space. The invariant
manifolds of these fixed points overlap, according to the continuous
translational symmetry of (\ref{1-1}), or due to the existence of
an integral of motion of (\ref{1-2}). 

There exist different possibilities to
modify the spatial differentials in (\ref{1-2}) into differences.
The most common way is to represent the differences in terms
of interaction forces between neighbouring particles $X_l$ and $X_{l-1}$:
\begin{equation}
-C(X_{l+1} + X_{l-1} - 2X_l)  + \frac{\partial V}{\partial \Phi}|
_{\Phi=X_l} = 0\;\;.
\label{1-3}      
\end{equation}
Here $X_l = \Phi(x=l)$, and $l$ is an integer (without loss of generality
the periodicity of the discrete chain is assumed to be equal to one).
Equation (\ref{1-3}) is a two-dimensional symplectic map, similar to the
standard map. In general the invariant manifolds of different
fixed points do not overlap anymore. Instead they generally 
intersect in heteroclinic points at nonzero angles. The iteration of
a heteroclinic point is again a heteroclinic point. One can then
consider different sequences of heteroclinic points (let us call them
heteroclinic orbits). All of these orbits will be exponentially
attracted to the two fixed points for sufficiently large absolute values
of $l$. Exactly two of these orbits correspond to kink solutions,
and are thus related to their counterparts of the differential
equation (\ref{1-2}). However these two orbits have different energies
(in contrast to the differential equation case). The energy difference
is called Peierls-Nabarro barrier.

Let us note that there exist also choices of the difference operator
such that the invariant manifolds still overlap \cite{sw94}. In that
nongeneric case static kink solutions exist, which can be positioned
at any place on the lattice. However, the difference operators
are rather unphysical, and we will not consider these nongeneric
cases here.

In the limit $C \rightarrow \infty$ the two kink-type heteroclinic
orbits of (\ref{1-3}) approach their counterparts of (\ref{1-2}). This
is due to the fact, that large values of $C$ imply slow variations
of these solutions as compared to the lattice spacing. Consequently
it is tempting to use a perturbation approach, which links the
kink solutions of (\ref{1-2}) with the adjoint solutions of (\ref{1-3}).
In this paper we will present a first-order perturbation calculation
for the heteroclinic orbits of (\ref{1-3}). In section II
the difference equation (\ref{1-3})
is transformed into a differential equation with spatially modulated
densities. This differential equation is analyzed in section III with
the help of separation into slow and fast variables, such that analytical
expressions for the kink solutions of (\ref{1-3}) (in first order perturbation
theory) are obtained. In section IV we apply our method to two
model cases and derive explicit expressions for the kink solutions.
Section V is devoted to a discussion of the calculation of the
Peierls-Nabarro energy.

\section{Reformulation of the problem}

Let us consider the following differential equation:
\begin{equation}
C\Phi_{,xx} - \rho (x) V'(\Phi) = 0 \;\;. \;\;\label{2-1}       
\end{equation}
Here $A_{,x}$ means (partial) derivative of $A$ with respect to $x$,
and $V'$ is the derivative of the potential $V(z)$.

If we choose $\rho(x)=1$, we obtain (\ref{1-2}). If we choose
\begin{equation}
\rho(x)=\sum_{l=-\infty}^{+\infty} \delta (x-l) \label{2-2}      
\end{equation}
we obtain (\ref{1-3}) \cite{suz88}. This is easy to see by the following
reasons. First we note that $\Phi_{,xx}(l < x < (l+1))=0$ or
$\Phi_{,x}(l< x < (l+1))=$ const. 
Thus it follows
\begin{equation}
\Phi(l+1) - \Phi(l) = \Phi_{,x}(l+0.5)\;\;.\label{2-3}         
\end{equation}
By integrating (\ref{2-1})
from $x=l-0.5$ to $x=l+0.5$ and using (\ref{2-2}) we obtain
\begin{equation}
C(\Phi_{,x}(l+0.5) - \Phi_{,x}(l-0.5) - V'(\Phi(l)) = 0 \;\;. \label{2-4}      
\end{equation}
Combining (\ref{2-3}) and (\ref{2-4}) we arrive at equation (\ref{1-3}),
where $\Phi(l)=X_l$. In other words, the field $\Phi(x)$ is given
by straight lines connecting its values at integer $x=l$, the 
field $\Phi_{,x}$ is given by a function with finite steps at
integer $x=l$ and constant elsewhere, and $\Phi_{,xx}$ is a sum
over delta-functions, with weights given by (\ref{2-1}) using (\ref{2-2}).

It is clear that one can 
make a continuous transition from (\ref{2-2}) to (\ref{2-3})
by varying $\rho(x)$ from $\rho(x)=1$ to
(\ref{2-2}).

We rewrite (\ref{2-2}) into
\begin{equation}
\rho(x)=\sum_{l=-\infty}^{+\infty} \delta (x-l) =
1 +2 \sum_{k=1}^{\infty} {\rm cos}(2\pi k x)\;\;. \label{2-5}       
\end{equation}
Thus we finally arrive at the following equation
\begin{equation}
C\Phi_{,xx} - (1 + 2\sum_{k=1}^{\infty} {\rm cos}(2\pi k x)) 
V'(\Phi) = 0 \;\;. \;\;\label{2-6}     
\end{equation}
Note that (\ref{2-6}) is still an exact reformulation of (\ref{1-3}).

\section{Perturbation approach}

Let us introduce new coordinates $x=\sqrt{C} T$ and $\Omega = 2 \pi \sqrt{C}$.
Then (\ref{2-6}) becomes
\begin{equation}
\Phi_{,TT} - (1 + 2\sum_{k=1}^{\infty} {\rm cos}(k\Omega T))
V'(\Phi) = 0 \;\;. \;\;\label{3-1}
\end{equation}
In the limit of large values of $C$ the cosine terms in (\ref{3-1})
rapidly oscillate due to the increase in $\Omega$. Thus we can apply standard
perturbation treatments using the separation of the field $\Phi$ into
slow $\Phi^{(s)}$ and fast $\xi_k$ parts \cite{LLI}:
\begin{equation}
\Phi = \Phi^{(s)} + \sum _{k=1}^{\infty} \xi_k \;\;. \label{3-2}    
\end{equation}
Inserting (\ref{3-2}) into (\ref{3-1}) and linearizing with respect to
the variables $\xi_k$ we obtain
\begin{equation}
\Phi^{(s)}_{,TT} + \sum_{k=1}^{\infty}\xi_{k,TT} = 
(1 + 2\sum_{k=1}^{\infty} {\rm cos}(k\Omega T))\left[
V'(\Phi^{(s)}) + V''(\Phi^{(s)})\sum _{k=1}^{\infty} \xi_k
\right] \;\;. \label{3-3}       
\end{equation}
For the fast variables the leading order contribution yields
\begin{eqnarray}
\xi_{k,TT} = 2 {\rm cos}(k\Omega T) V'(\Phi^{(s)})\;\;, \label{3-4} \\
\xi_k=-\frac{2}{k^2 \Omega^2} {\rm cos}(k \Omega T) V'(\Phi^{(s)})\;\;.
\label{3-5}      
\end{eqnarray}
Averaging (\ref{3-3}) over the periods of oscillation of the fast variables
and using (\ref{3-5}) and $\sum_{k=1}^{\infty}1/k^2 = \pi^2 / 6$ it follows
\begin{eqnarray}
\Phi^{(s)}_{,TT} = V_{\rm eff}'(\Phi^{(s)})\;\;,\label{3-6} \\
V_{\rm eff}(z) = V(z) - \frac{1}{24C}\left( V'(z)\right)^2 \;\;. 
\label{3-7}         
\end{eqnarray}
Note that equation (\ref{3-6}) is a simple differential equation, which
will be integrated for two examples in the following section.

Since we are interested in the solution of (\ref{3-1}) at integer points,
the arguments $k\Omega T= 2\pi n$ with $n$ being an integer in (\ref{3-5}).
The final solution of (\ref{3-1}) to first order in $1/C$ is then given
by
\begin{equation}
\Phi(l) = \Phi^{(s)}(l) - \frac{1}{12C}V'(\Phi^{(s)}) \;\;. \label{3-8}
\end{equation}
Actually (\ref{3-8}) contains also (incomplete) terms of order
$1/C^2$. One can expand the
solution in powers of $1/C$ and extract the first order term
after solving along the given path.

\section{Two examples}

\subsection{sine Gordon case}

Let us consider 
\begin{equation}
V(z)=1-{\rm cos}(z)\;\;. \label{4-1}
\end{equation}
The kink solution of (\ref{1-2}) is given by
\begin{equation}
\Phi^{(c)}(x+ \alpha)=4 {\rm arctan}({\rm e}^{x\sqrt{C}})\;\;. \label{4-2}      
\end{equation}
Let us consider the slow part $\Phi^{(s)}$ of the first-order perturbation.
The effective potential (\ref{3-7}) is given by
\begin{equation}
V'_{\rm eff}(z) = {\rm sin}(z) - \frac{1}{24C}{\rm sin}(2z)\;\;. \label{4-3}   
\end{equation}
Consequently $\Phi^{(s)}$ is the solution of the double sine Gordon
equation and can be found in \cite{cgm83} (note that there is an error
in equation (3.7) of \cite{cgm83} - the sign of the power $-1/2$ has to be
changed to $+1/2$) or can be simply calculated by integration:
\begin{eqnarray}
\Phi^{(s)}(x+\alpha) = 2\pi - 2{\rm arctan}\left[ \sqrt{1-\frac{1}{12C}}
{\rm cosech} \left( \sqrt{1-\frac{1}{12C}} \frac{x}{\sqrt{C}} \right)  
 \right] \;\; , \;\; x\geq 0 \;\;, \label{4-4} \\
\Phi^{(s)}(x+\alpha) = - 2{\rm arctan}\left[ \sqrt{1-\frac{1}{12C}}
{\rm cosech} \left( \sqrt{1-\frac{1}{12C}} \frac{x}{\sqrt{C}} \right)
 \right] \;\;, \;\; x\leq 0 \;\;. \label{4-5} 
\end{eqnarray}
Here $\alpha$ is an integration constant.
Using equation (\ref{3-8}) and expanding in $1/C$ we finally obtain
the following first-order perturbation correction for the 
discrete sine Gordon chain:
\begin{equation}
\Phi(l) = \Phi^c(x+\alpha) + \frac{1}{6C}{\rm sech}\left( \frac{x+\alpha}
{\sqrt{C}}\right) \left[ 2 {\rm tanh}\left( \frac{x+\alpha}{\sqrt{C}}\right)
- \frac{x+\alpha}{\sqrt{C}} \right]\;\;. \label{4-6}      
\end{equation}
Since the invariant manifolds of the two relevant fixed points of
(\ref{1-3}) do not overlap, but only intersect at finite angles, we
have to choose the right values of $\alpha$. Clearly they are 
$\alpha=0$ and $\alpha=0.5$, which correspond to a kink centered
on a lattice site and between two lattice sites respectively. These
two possible kink solutions are known to exist for the map (\ref{1-3})
\cite{pr82}-\cite{tm92}.

In order to test our result we compute the exact kink solutions of
(\ref{1-3}) with (\ref{4-1}) for different values of $C$. We use 
the steepest gradient method (minimization of the potential energy) 
and work in quadruple precision. The result will be denoted as
$X_l$. The deviations $d_l$ from its adjoint solution (\ref{4-2}) of
(\ref{1-2}) is then given by $d_l=X_l - \Phi^c(l)$. The perturbation
approach yields $\phi(l) = \Phi(l) - \Phi^c(l)$ and is defined by
the second term on the rhs of (\ref{4-6}). In Fig.1 we plot
$d_l$ and $\phi_l$ for $C=10$ for both kink solutions ($\alpha=0$
and $\alpha=0.5$). Clearly the perturbation result
fits well to the exact one. In order to be more precise, we
calculate the normalized squared deviation $\Delta$ of the perturbation result
from the exact one:
\begin{equation}
\Delta = \frac{\sum_{l=-\infty}^{+\infty} (d_l - \phi_l)^2}
{\sum_{l=-\infty}^{+\infty} d_l^2} \;\;. \label{4-7}   
\end{equation}
Now we can evaluate $\Delta$ for different values of $C$ and see, whether
it is monotonously decreasing with increasing $C$. The results for
both kink solutions are shown in Fig.2. No doubt the perturbation theory
gives the correct first order result.

\subsection{$\Phi^4$ case}

The second example is given by 
\begin{equation}
V(z) = \frac{1}{4}(z^2-1)^2\;\;. \label{4-8}      
\end{equation}
The kink solution of (\ref{1-2}) is given by
\begin{equation}
\Phi^c(x+\alpha)={\rm tanh}\left(\frac{x}{\sqrt{2C}}\right)\;\;. \label{4-9}    
\end{equation}
The effective potential (\ref{3-7}) reads
\begin{equation}
V_{\rm eff}(z) = \frac{1}{4}(z^2 - 1)^2\left[ 1 - \frac{z^2}{6C} \right]
\;\;. \label{4-10}       
\end{equation}
Thus the slow part $\Phi^{(s)}$ is the solution of the $\Phi^6$ differential
equation. It can be easily integrated using \cite{gr94}:
\begin{equation}
\Phi^{(s)}(x+\alpha) = \frac{{\rm tanh}\left( \sqrt{1-\frac{1}{6C}} 
\frac{x}{\sqrt{2C}}     \right) } 
{ \sqrt{ 1-\frac{1}{6C}{\rm sech}^2
\left( \sqrt{1-\frac{1}{6C}}\frac{x}{\sqrt{2C}}          \right) } }        
\;\;. \label{4-11}
\end{equation}
Using (3-8) we finally obtain the first order perturbation result
\begin{equation}
\Phi(l)= \Phi^{(s)}(x+\alpha) - \frac{1}{12C}\Phi^{(s)}(x+\alpha)
(\Phi^{(s)}(x+\alpha)\Phi^{(s)}(x+\alpha) - 1)\;\;. \label{4-12}       
\end{equation}
As in the sine Gordon case we calculate $d_l$ and $\phi_l$ and plot
the results for $C=15$ in Fig.3. The normalized deviation $\Delta(C)$
is plotted in Fig.4. Clearly the perturbation theory gives the correct
result.

\section{The Peierls-Nabarro barrier problem}

Considering the success of the presented perturbation approach
with respect to the kink solutions, it is tempting to 
use this result for calculating the Peierls-Nabarro barrier $E_{PN}$
which is given by the energy difference of the two different
kink solutions. However as it was shown in \cite{fw1}, one has
to expect that the leading order asymptotics of $E_{PN}$ 
contains contributions from all orders of the perturbation
series for the kink solutions
for large
values of $C$. 
This is already clear by noting
that the zero-order result (i.e., replacing the exact kink solution
of the lattice by its counterpart of the adjoint differential equation)
yields a nonzero $E_{PN}^{(0)}$. As shown in \cite{fw1}, these contributions
are not enough to fit the exact numbers. Clearly at least the
first-order perturbation result for the discrete kink has to be taken
into account (yielding $E_{PN}^{(1)}$). 
But then it is clear that contributions have to
expected throughout all higher orders of perturbation theory.

Let us introduce $R^{(1)}= E_{PN}^{(1)}/E_{PN}$ and 
$R^{(0)}= E_{PN}^{(0)}/E_{PN}$, which measure the ratio of
the first order energy difference (zero order respectively)
over the exact one. In Fig.5 these results are plotted for
the two examples considered in the previous section. Clearly 
$R^{(1)}$ is much closer to unity than $R^{(0)}$, but still there
exist discrepancies, which even grow with increasing $C$. This
circumstance implies, that the contributions from higher orders
of the presented perturbation theory in $E_{PN}$ 
gain more weight with increasing
$C$. Consequently the first order perturbation scheme, which works
as better for the kink solution as larger $C$ is, works as better
for the Peierls-Nabarro energy as {\sl smaller} $C$ is. So there
is an intermediate range of values of $C$, where the first order
perturbation theory can be satisfactory applied to calculate
both kink shape and energy difference.

Since the calculation of the energy difference using our derived
first-order perturbation results is still a matter of computing
sums, it is not very practically either. Consequently the value
of the presented approach is clearly in calculating analytically
the first-order corrections to the kink shape (rather than the
energy difference). 

There exists a theoretical approach to obtain the Peierls-Nabarro energy
by using properties of the splitting angle of the map (\ref{1-3})
\cite{lst89}.
The main point is, that the asymptotic dependence of $E_{PN}$ on
$C$ for large $C$ can be substituted by a form which contains
higher orders. Let us briefly explain this using the sine Gordon
chain as an example. The asymptotic $E_{PN}(C)$-dependence for large $C$
is given by \cite{fw1}
\begin{equation}
E_{PN} = Z C {\rm e}^{-\pi^2 \sqrt{C}}\left( 1 + O(\frac{1}{C}) \right)
+ O\left( {\rm e}^{2\pi^2 \sqrt{C}}\right) \;\;. \label{5-1}     
\end{equation}
Here the constant $Z$ is a prefactor, and usually all perturbation
approaches are aimed to fix this constant. 

To give an example, let
us start with the zero-order perturbation with respect to the kink shape,
i.e., let us take the continuum kink solution (\ref{4-2}) and calculate
the energy difference $E_{PN}^c$ and then $Z_c$. 
Naive approaches of this particular task have
in addition expanded the differences in the energy expression $(X_l-X_{l-1})$
into series over differentials (see e.g. \cite{vlp81}. 
Then noting that the continuum kink shape
causes higher order derivatives to decay faster with increasing $C$
these higher derivatives were dropped in the mentioned naive approach.
In the energy expression that means taking into account only squared
second derivatives, and in the collective coordinate frame (equations of
motion for the kink) it would mean taking into account only the fourth order
derivative of the continuum kink shape \cite{fw1}. 
As it was shown in \cite{fw1}, the
dropping of higher order derivatives is wrong, since all of them contribute
in leading order to $Z_c$. The correct result for $Z_c$ gives
\cite{fw1} 
\begin{equation}
Z_c = 2^5\left[ \frac{2}{3}\pi^2 - 1\right] +
\sum_{m=2}^{\infty}(-1)^m\frac{2^{2m+4}\pi^{2m+1}}{m(2m+1)!}\approx 237.82...
\;\;. \label{5-2}      
\end{equation}
To be on the save side we plot in Fig.6 the ratio 
\begin{equation}
\tilde{Z}_c=E_{PN}^c/\left( C {\rm e}^{-\pi^2 \sqrt{C}}\right)\;\;. \label{5-3}
\end{equation}
In the limit $1/C \rightarrow 0$ $\tilde{Z}_c$ should approach $Z_c$.
Clearly this is the case, demonstrating unambiguously that {\sl higher
order derivatives in the continuum kink shape contribute in leading order
to $Z_c$}. To be precise the naive approach (truncating higher
order derivatives) yields the result \cite{vlp81}
\[
64\pi^2 \approx 631.65...
\]
To compare with the exact result we plot $\tilde{Z}$ for the exact 
Peierls-Nabarro barrier in Fig.6 . Clearly the zero order perturbation
fails by more than 60 percent. 

Finally we plot in Fig.6 $\tilde{Z}$ for the first-order perturbation
result derived in this paper. We are about 20 percent off the asymptotic
behaviour. Note that both the exact and first order result for $\tilde{Z}$
are rather strongly varying with $1/C$, so that the true asyptotic
dependence (\ref{5-1}) is practically never reached. Here the mentioned
approach through the splitting angle \cite{lst89} 
has clear advantages. The map (\ref{1-3})
is linearized around one of its hyperbolic fixed points. The eigenvalues
are given by
\begin{equation}
\lambda_{\pm} = 1+\frac{1}{2C} \pm \sqrt{\frac{1}{C}+ \frac{1}{4C^2}}\;\;,
\;\;Z_+Z_-=1\;\;. \label{5-4}     
\end{equation}
If we replace the exponent $(-\pi^2 \sqrt{C})$ by 
$(-\pi^2 /{\rm ln}(\lambda_+))$ then we can write
\begin{equation}
E_{PN} = \bar{Z} C {\rm e}^{-\pi^2 /{\rm ln}(\lambda_+)}
\;\;. \label{5-5}
\end{equation}
$\bar{Z}$ is plotted in Fig.6 for the exact value of $E_{PN}$. Clearly
the exact result is described by the dependence (\ref{5-5}) very good even
down to small values of $C$, so that: i) (\ref{5-5}) is describing the true
$E_{PN}(C)$ dependence much better than (\ref{5-1}); ii) (\ref{5-5}) is an
easy to use expression; iii) the constant $Z$ can be very precisely determined
by analyzing the weak dependence of $\bar{Z}$ on $1/C$. This yields
\cite{lst89}
\[
Z \approx 712.26...\;\;.
\]
Having this number, one can determine the exact Peierls-Nabarro energy
of the sine Gordon chain using (\ref{5-5}) with very good precision, and
down to values $C \sim 1$.

\section{Concluding remarks}

We have derived first order corrections to the kink shape of a discrete
chain. We used the methods of slow and fast variables. The resulting
differential equations can be integrated explicitely, as demonstrated
for two examples.
Note that the presented method can be generalized to the case
of anharmonic interactions  as well as to time-dependent solutions.
The generalization of (\ref{2-1}) gives
\begin{equation}
\sum_{l}\delta(x-l) \Phi_{,tt} - \frac{\partial ^2 W}{\partial y^2}
|_{y=\Phi_{,x}}\Phi_{,xx} + \sum_{l}\delta(x-l) V'(\Phi) = 0 \;\;.     
\label{6-1}
\end{equation}
Here $W(y)$ denotes the nearest neighbour interaction on the discrete
chain ($y=X_l-X_{l-1}$), 
which could well be anharmonic. In the examples considered
above we used only harmonic interactions
\[
W(y)=\frac{1}{2}C y^2 
\]
so that the second derivative in (\ref{6-1}) simply yields $C$. 
\\
\\
\\
\\
Acknowledgements
\\
\\
We thank S. Aubry, P. Fulde, R. S. MacKay, C. Simo, O. Usatenko 
and C. R. Willis  
for valuable discussions,
and C. Simo for providing
us with numerical details on the splitting angle approach.
K.K. is grateful to the Max-Planck-Institut f\"ur Physik komplexer
Systeme for the kind support.

\newpage

FIGURE CAPTIONS
\\
\\
\\
Fig.1
\\
\\
Discrete kink shape deviations from the kink solution of the adjoint
differential equation versus lattice site $l$ for the sine Gordon
chain with $C=10$. 
\\
Cirlces - exact
result $d_l$; 
\\
crosses - first order perturbation result $\phi_l$.
\\
a) $\alpha=0.5$; b) $\alpha=0$.
\\
\\
\\
Fig.2
\\
\\
The normalized deviation $\Delta$ of the first order perturbation result
$\phi_l$ from the exact $d_l$ versus $C$. 
\\
Circles - $\alpha=0.5$,
crosses - $\alpha=0.5$.
\\
\\
\\
Fig.3
\\
\\
Same as in Fig.1 but for the $\Phi^4$ chain and $C=15$.
\\
\\
\\
Fig.4
\\
\\
Same as in Fig.2 but for the $\Phi^4$ chain. 
\\
\\
\\
Fig.5
\\
\\
The ratio $R$ of the approximated $E_{PN}$ over the exact one
as a function of $C$. 
\\
Open circles - zero order result for sine Gordon
chain; 
\\
filled circles - first order perturbation result for sine Gordon
chain; 
\\
open squares - zero order result for $\Phi^4$ chain; 
\\
filled squares -
first order perturbation result for $\Phi^4$ chain.
\\
Lines are guides to the eye.
\\
\\
\\
Fig.6
\\
\\
$\tilde{Z}$ versus $1/C$ as a test for the asymptotic behaviour
of $E_{PN}$ for the sine Gordon chain (see text). 
\\
Open circles - zero order result;
\\
filled circles - first order result;
\\
open squares - exact result;
\\
filled squares - $\bar{Z}$ for the exact result (see text).
\\
Solid lines are guides to the eye. 
\\
Dashed line - predicted value $Z\approx 712.26...$ for the exact
result. Clearly $\bar{Z}$ is approaching this value for $1/C\rightarrow
0$. Also the exact result for $\tilde{Z}$ (open squares) is approaching
the dashed line in the same limit.
\\
Dotted line - $Z=64\pi^2 \approx 631.65...$ (see text).

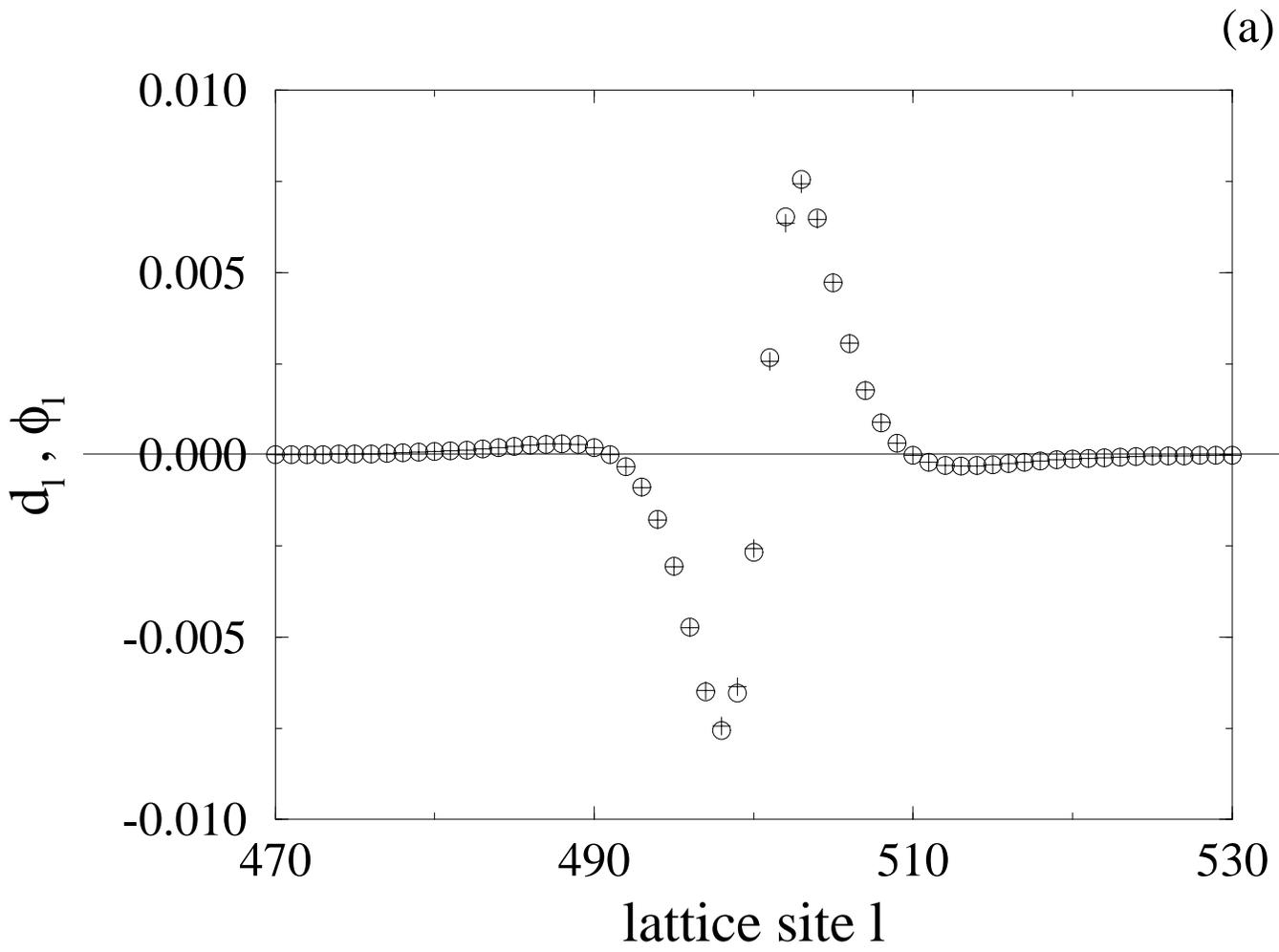
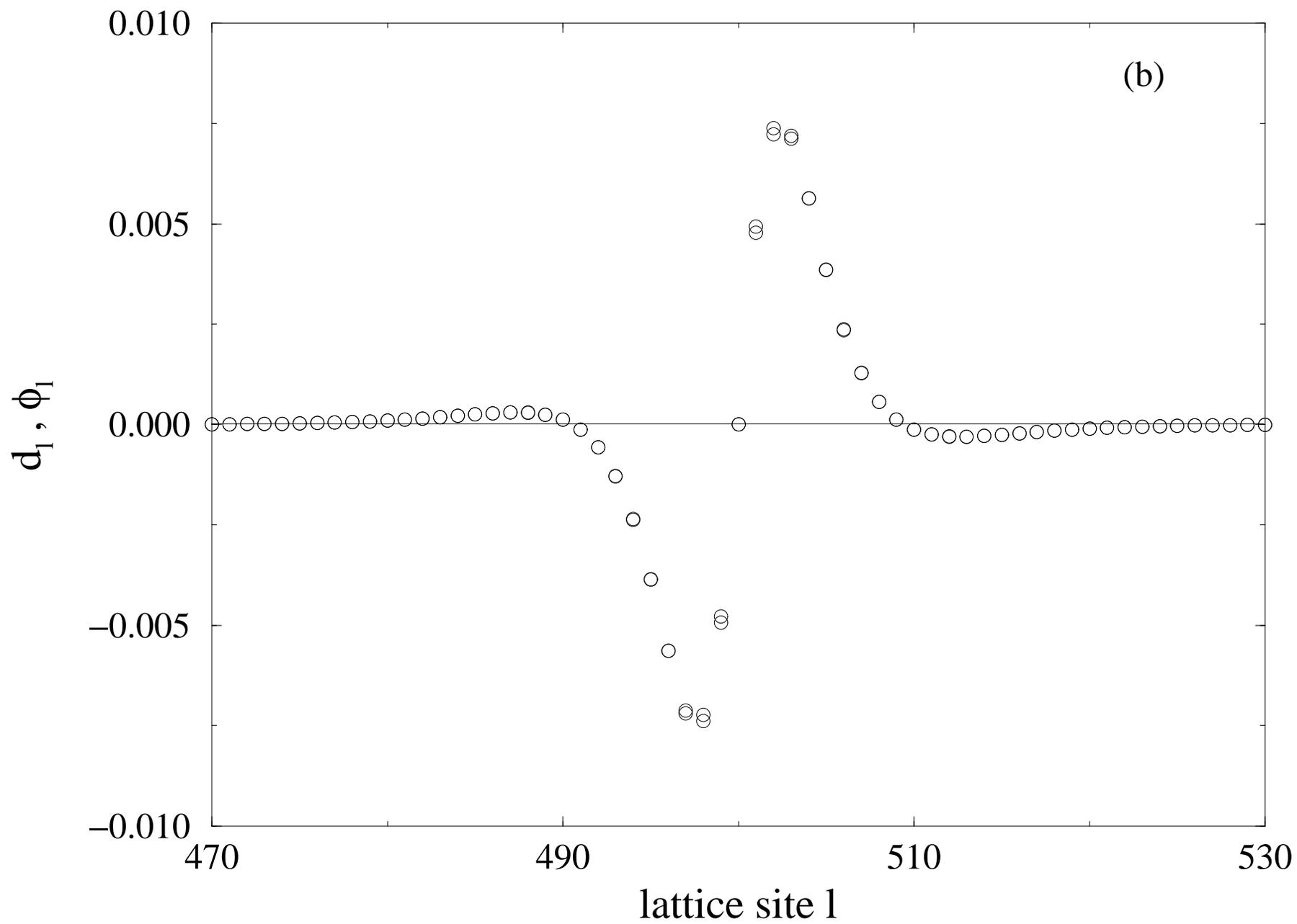
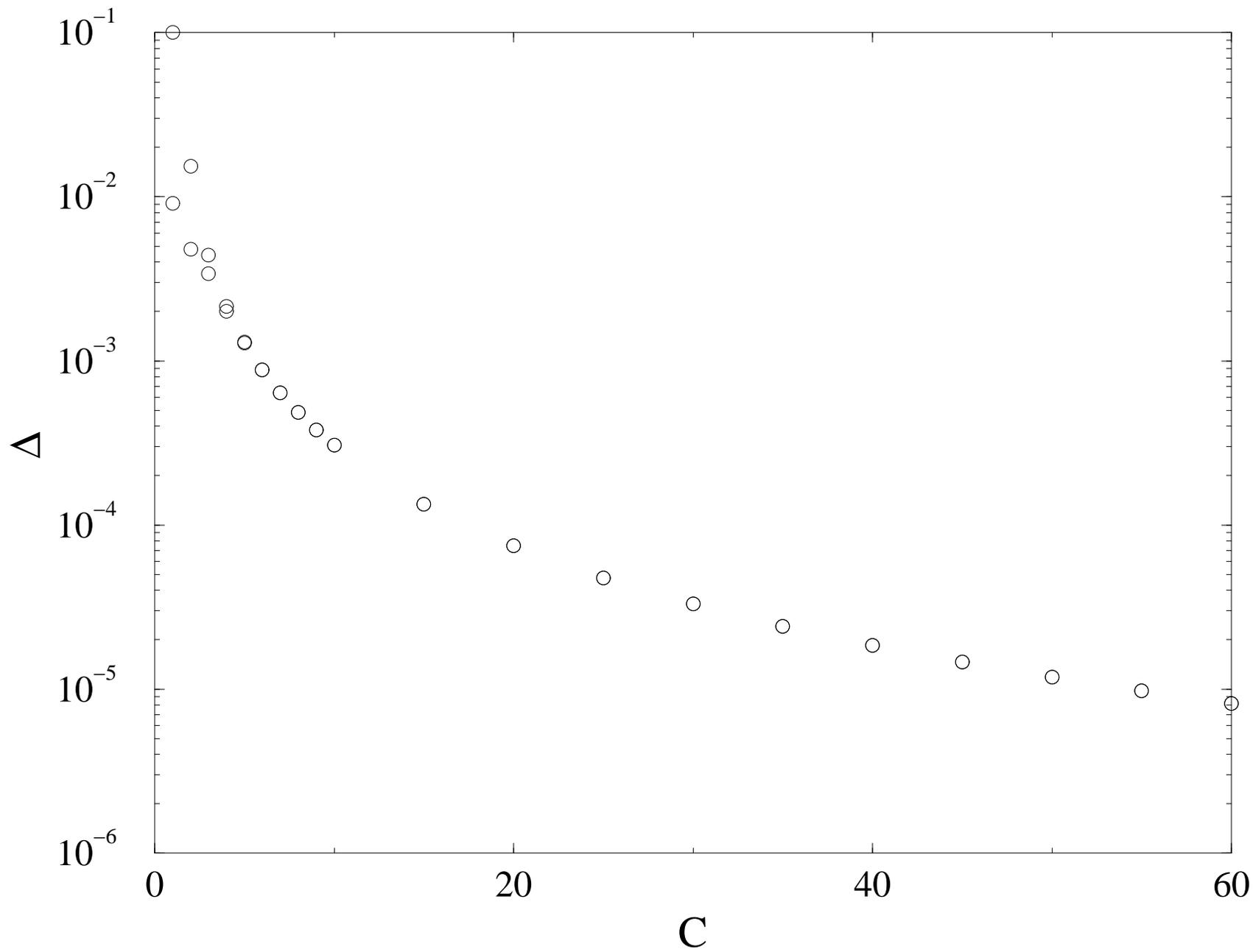
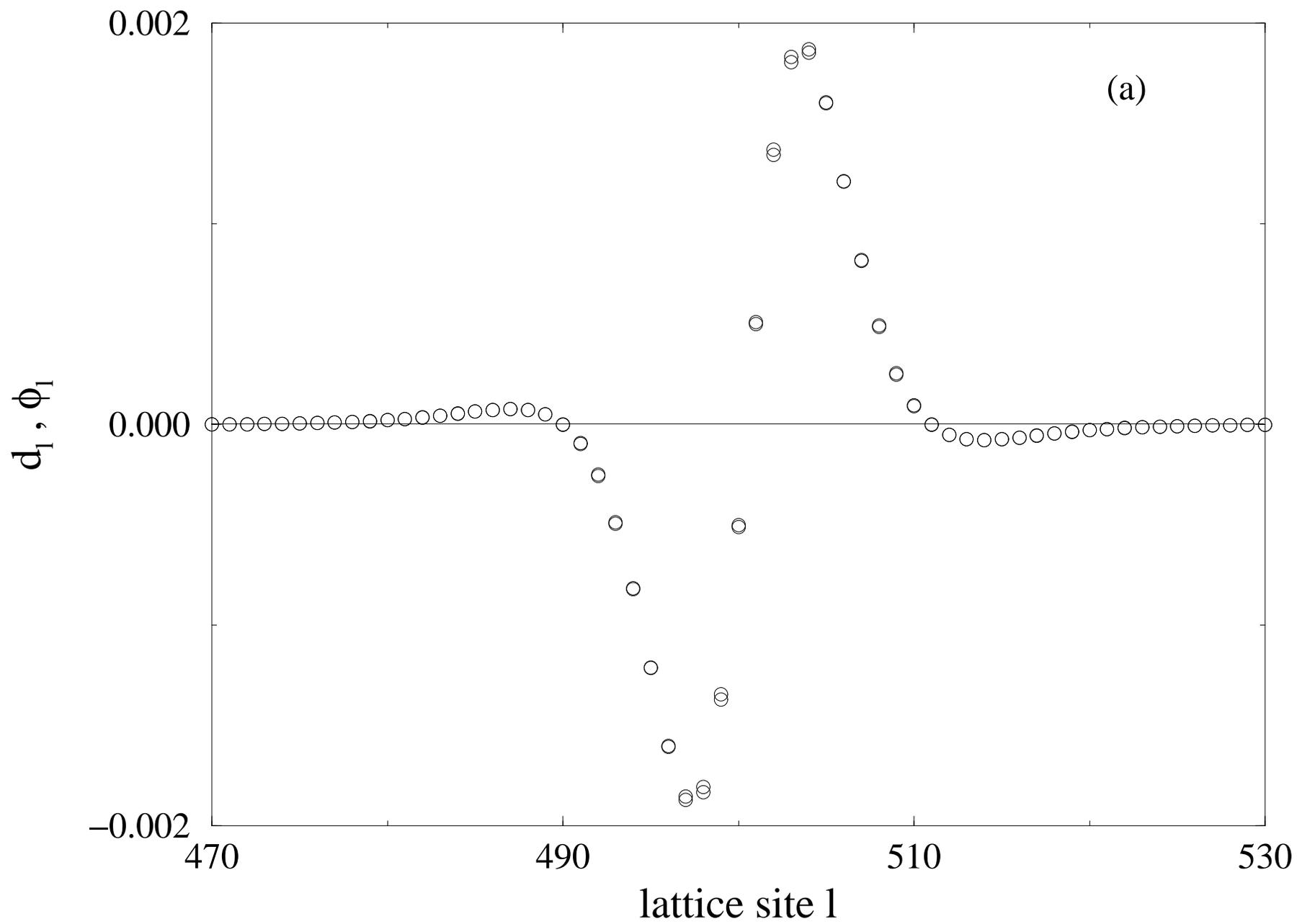
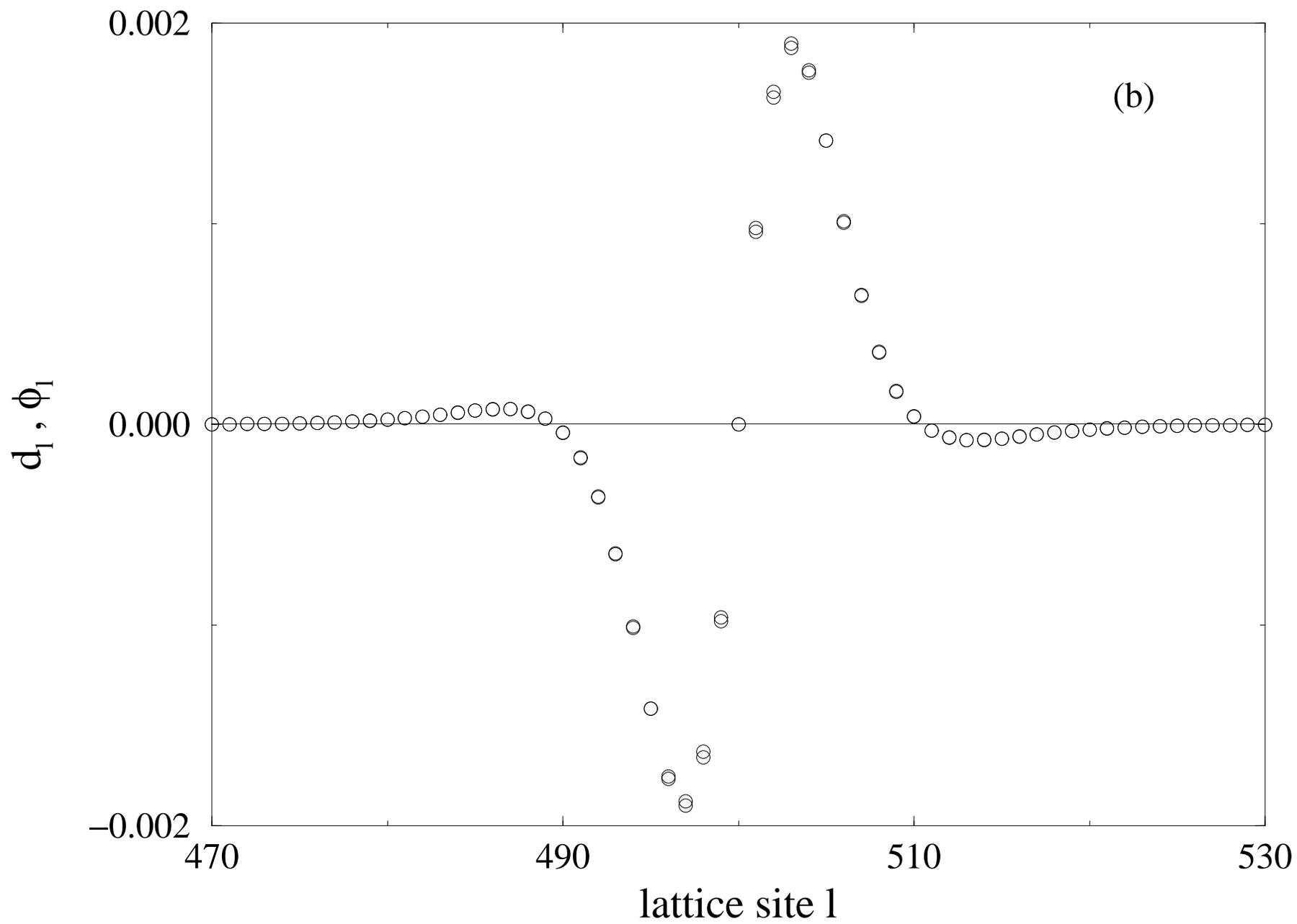
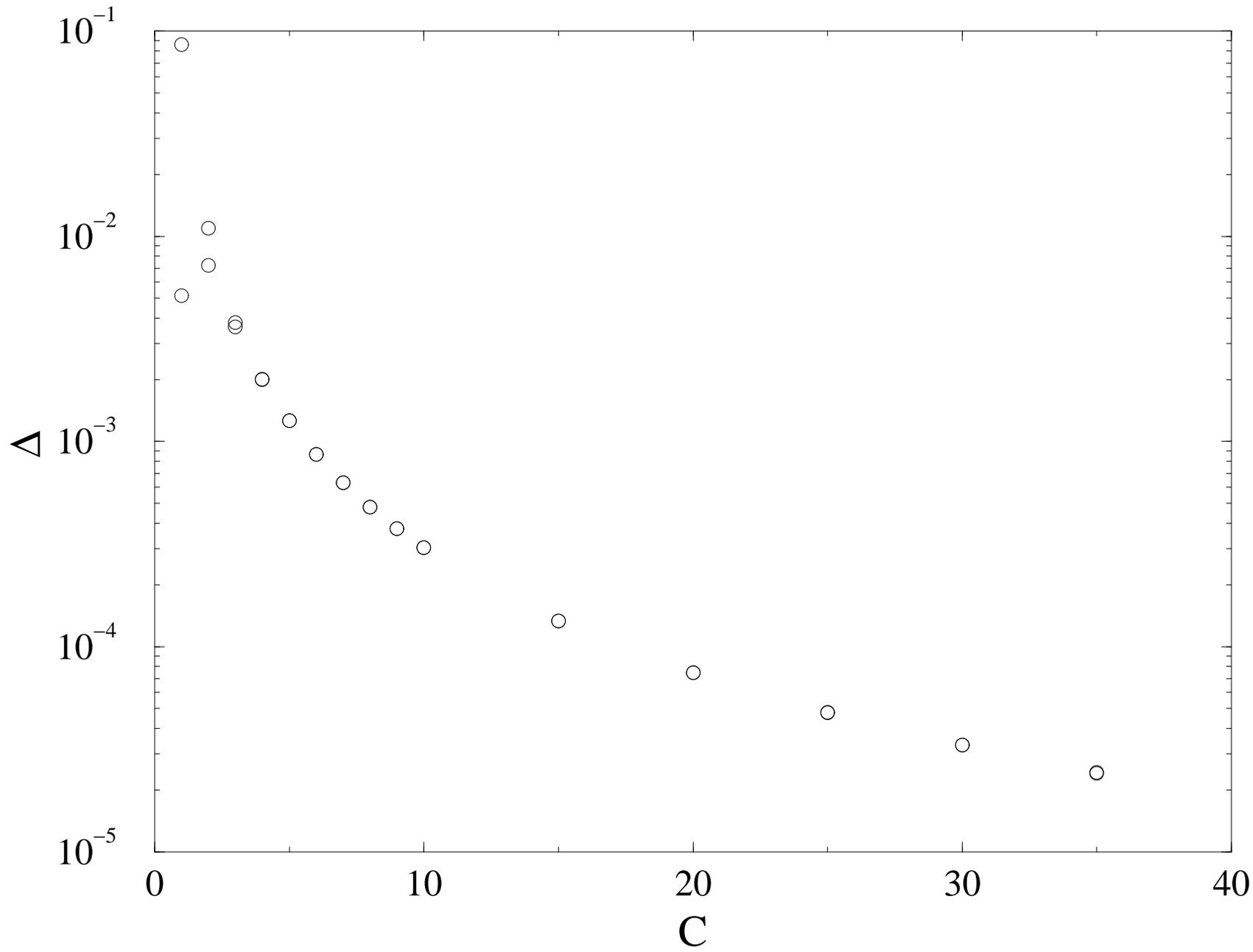
\end{document}